\documentclass[fleqn,10pt]{wlscirep}
\usepackage[utf8]{inputenc}
\usepackage[T1]{fontenc}
\usepackage{xcolor}
\usepackage{amsmath}
\usepackage{amssymb}
\usepackage{graphicx}
\usepackage{fullpage}
\usepackage{setspace}
\usepackage{tabularx} 
\usepackage{booktabs} 

\title{Redefining Fitness: Inference, Information and Phase Transitions in Evolutionary Dynamics}

\author[1,2,*]{Lu\'is M. A Bettencourt}
\author[1]{Brandon J. Grandison}
\author[2,3,4]{Jordan T. Kemp}
\affil[1]{University of Chicago, Department of Ecology and Evolution, Chicago IL, 60637, USA}
\affil[2]{Santa Fe Institute, Santa Fe NM, 87501, USA}
\affil[3]{University of Oxford, Institute for New Economic Thinking at the Oxford Martin School, OX1 3UQ, UK}
\affil[4]{University of Oxford, School of Geography and the Environment, OX1 3QY, UK}

\affil[*]{corresponding author(s): Lu\'is Bettencourt (bettencourt@uchicago.edu)}


\begin{abstract}
Evolution is the adaptation of populations to their environment expressed through the concept of fitness.  Darwin did not define fitness but described evolution as the higher prevalence of lineages with advantages in survival and reproduction in changing environments, an implicitly statistical and relational notion. As evolutionary dynamics became more quantitative, however, fitness acquired a narrower meaning of relative reproductive success. Crucially, this narrower definition suffers from three fundamental difficulties, known as the circularity, mismatch, and prediction problems.  We show that interpreting evolutionary dynamics in terms of inference resolves these three problems while also creating new productive analytical tools. This shift redefines fitness via a Bayesian likelihood, a predictive probability of the environment specific to each type. We show that averaging the growth rate over environmental histories connects selection to information as types with better environmental models are amplified. It follows that long-run evolutionary dynamics maximizes the mutual information between population structure and environmental statistics, establishing information maximization as the governing principle of natural selection. We illustrate this approach in several population dynamics problems including task switching, evolutionary games, and selection in group-structured populations. In each case, we derive phase diagrams as functions of environmental statistics and Hamilton-type rules for the emergence of cooperation, while also demonstrating the generality of the approach.
\end{abstract}

\begin{document}

\flushbottom
\maketitle

\thispagestyle{empty}

\section*{Introduction}
Our understanding of evolution rests on the concept of fitness.  Darwin's original formulation of evolution by natural selection never quite defined fitness~\cite{darwin_origin_1859}. His description was purely qualitative, invoking  environment-specific advantages in survival and  reproduction in terms of a "struggle for existence" under  ever-varying conditions~\cite{darwin_origin_1859}. In this sense, the adaptations underpinning evolution have always been understood to be implicitly statistical and relational to the environment.  However, as the fields of ecology and evolution became operationalized mathematically, fitness took a more precise but narrower meaning, losing its probabilistic environmental dependence. In the standard mathematical formulations we use today~\cite{crow1970introduction,Frank2011}, fitness is defined as relative reproductive success — the ratio of numbers of offspring to parents of a given type.  Evolutionary adaptation is then identified with a type's relative amplification across time (generations), over and above random effects such as drift.  This framework underlies our general quantitative understanding of ecology and evolution from population genetics to evolutionary game theory~\cite{maynard_smith_evolution_1982,traulsen2023future}, the Price equation~\cite{frank_natural_2012}, and models of structured populations~\cite{Ohtsuki2006,szathmary_major_1995}.

However, it has long been recognized that defining fitness as relative reproductive success is both theoretically incomplete and 
practically limiting~\cite{sep-fitness,Millstein2016}. First, defining  fitness in terms of reproductive success is logically 
circular~\cite{sep-fitness,Millstein2016}: a type that reproduced  more is called fitter; a fitter type reproduces more. This explains 
neither why nor how a type is well adapted to its environment nor does it predict when the same type will be favored 
in the future. Second, this issue is compounded by the so-called \emph{mismatch}  problem~\cite{Millstein2016}: if fitness equals reproductive 
output, then a well-camouflaged butterfly that happens to be eaten  before reproducing has zero fitness, while a conspicuous one that 
survives by chance is deemed fitter. Thus, what we mean by fitness is not the outcome of a 
particular instance but an underlying capacity and its associated probability. This is the core  insight of the {\it propensity interpretation} of 
fitness~\cite{Mills1979,Brandon1990,Sober1984,Millstein2016}:  fitness is a causal but probabilistic disposition of a phenotype grounded in its 
physical traits relative to a specific environment. 

Mathematically, these general considerations about propensity, prediction, and statistical dependence on the environment are naturally formalized in terms of conditional probabilities \cite{Millstein2016,Okasha2013,Czegel2022}. The probability of environmental states, $e$, given types $s$, $p(e|s)$ turns out to be a likelihood function, which changes the prevalence of each type given events in the environment. This is nothing more than a Bayesian update to the relative frequency of each type, a probabilistic model of types evaluated against experienced environmental events~\cite{mackay_information_2003,csillag_bayes_darwin_2025}.  Here we show that this identification redefines fitness and resolves the circularity, mismatch and prediction problems, along with introducing methods to solve hitherto difficult evolutionary problems. By separating the probabilistic model of fitness from its actual outcome, and by clarifying what adaptation means in terms of the relative states of the population and the environment, this approach also closes the conceptual gap between ecology and evolution as a theory of metabolism and information and approaches in other disciplines from statistical physics to neuroscience and from data sciences to social behavior~\cite{bergstrom_shannon_2004,frank_natural_2012,adami2004information,adami_use_2012}.

We start by recognizing that the formal connection between natural selection and probabilistic inference has a long history. Working 
within population genetics in the 1960s, Kimura~\cite{kimura_natural_1961}  first noted that natural selection acts as a process of  accumulating genetic information, establishing an early link  between selection and information theory. In analogy with statistical  physics, Iwasa~\cite{iwasa_free_1988} introduced a ``free 
fitness'' quantity --- analogous to a free energy --- and showed that it increases monotonically under selection. Barton and 
colleagues~\cite{barton_statistical_2009,bodova_general_2016} 
later developed maximum-entropy methods to describe the dynamics 
of quantitative traits under selection, mutation, and drift.  From  the perspective of information theory and evolutionary biology, Bergstrom and 
Lachmann~\cite{bergstrom_shannon_2004} and  Frank~\cite{frank2009natural} showed that selection 
maximizes Fisher and Shannon information. 

The most direct antecedents of  the present work are recent proofs that the replicator equation of population genetics under 
pure selection maps exactly to Bayes' theorem of statistics. In this mapping, as already anticipated, fitness is written in terms of a likelihood function, type frequencies are priors, and mean fitness  is the normalization~\cite{harper2009replicator,Okasha2013,Czegel2022,czegel_multilevel_2019}, see below. 
Specifically, Harper~\cite{harper2009replicator}, to our knowledge, was the first to  establish this mapping 
mathematically. Okasha~\cite{Okasha2013} examined it from the  philosophy of biology; and Cz\'egel and 
colleagues~\cite{Czegel2022,czegel_multilevel_2019} developed it  as the computational basis of general evolutionary theory (beyond biology) and multilevel 
selection, followed by Campbell~\cite{Campbell2016} who proposed  it as the principle underlying universal Darwinism on any 
genetic or cultural substrate.

However, while suggestive, the mapping of evolutionary dynamics to statistical inference has remained mostly a formal result, yet to be shown to be productive. To address this issue, we operationalize this mathematical equivalence here as a constructive approach for solving concrete classes of problems in evolutionary theory.  We show that writing fitness explicitly in terms of a Bayesian likelihood has three productive consequences beyond the formal identification. First, it provides a direct recipe for building fitness models: any conditional probability of environmental states given a phenotype defines a valid fitness function, making the relationship between phenotypes and selective environments explicit and measurable. Second, because statistically independent fitness contributions multiply while their logarithms add, the average log-fitness (growth rate) becomes the mutual information between population structure and environmental statistics~\cite{tkavcik2016information,rivoire2011value}. Information maximization then emerges as the governing principle of natural selection in variable environments. Third, because most probability functions are exponential, their logarithms are simpler "energy" functions\cite{mackay_information_2003}, often in the form of polynomials in the type and environmental states. We show that expressing log-fitness as a polynomial in binary phenotypic variables reveals evolutionary phase transitions as sign changes in polynomial coefficients driven by environmental averages.  This yields a general analytical framework for identifying phase transitions in evolutionary dynamics, even without numerical simulation.

We illustrate these tools through three applications: task selection in stochastic environments, evolutionary games (including the Prisoner's Dilemma and mixed games in variable environments), and selection in group-structured populations. We show how the log-fitness framework yields phase diagrams for evolutionary outcomes in each case and recovers and extends Hamilton-type rules for the emergence of cooperation in a unified way. Throughout, we work with selection among asexual haploid replicators in discrete time — the setting where the Bayesian mapping is exact — and discuss where other evolutionary forces such as drift, mutation, and recombination modify these conclusions.

\section*{Results}
We now show how redefining fitness via concepts of inference quantifies biological adaptation to probabilistic environments and how this identification resolves the problems raised in the Introduction. 

\subsection*{Population replicators and growth dynamics}
We define a structured population evolving under replicator dynamics. The population is made up of different types, $s$, such that the total population size, $N=\sum_s N_s$, where $N_s$ is the population of each type at a given time.  The temporal evolution follows the standard replicator equation for a haploid population in discrete time $\tau$, $ N_s(t) = w_s N_s (t-\tau ),$
so that
\begin{eqnarray}
N(t)=\sum_s N_s (t) = \sum_s w_s N_s(t-\tau) = w N(t-\tau),  
\label{eq:fitness}
\end{eqnarray} 
where $w_s\geq 0$ is the type's fitness. The type $s$ will grow if $w_s>1$ and decrease for $w_s<1$. This leads to the growth equation for the total population with average fitness $w=\sum_s w_s f_s $. Here, $f_s = \frac{N_s}{N}$ is the fraction of the population of type $s$. The total population will grow if $w> 1$.  The growth rate, $\gamma_s$ (Malthusian parameter), is related to fitness by exponentiation $w_s=e^{\gamma_s \tau}\simeq 1+ \gamma_s \tau $, and, correspondingly $\gamma_s = \frac{1}{\tau} \ln w_s \simeq \frac{w_s -1}{\tau}$. These relationships apply in continuous time while the approximations are first-order series expansions. In the following, we set $\tau=1$ for simplicity.

\begin{figure*}
  \centering
  \includegraphics[width=\linewidth,scale=0.6]{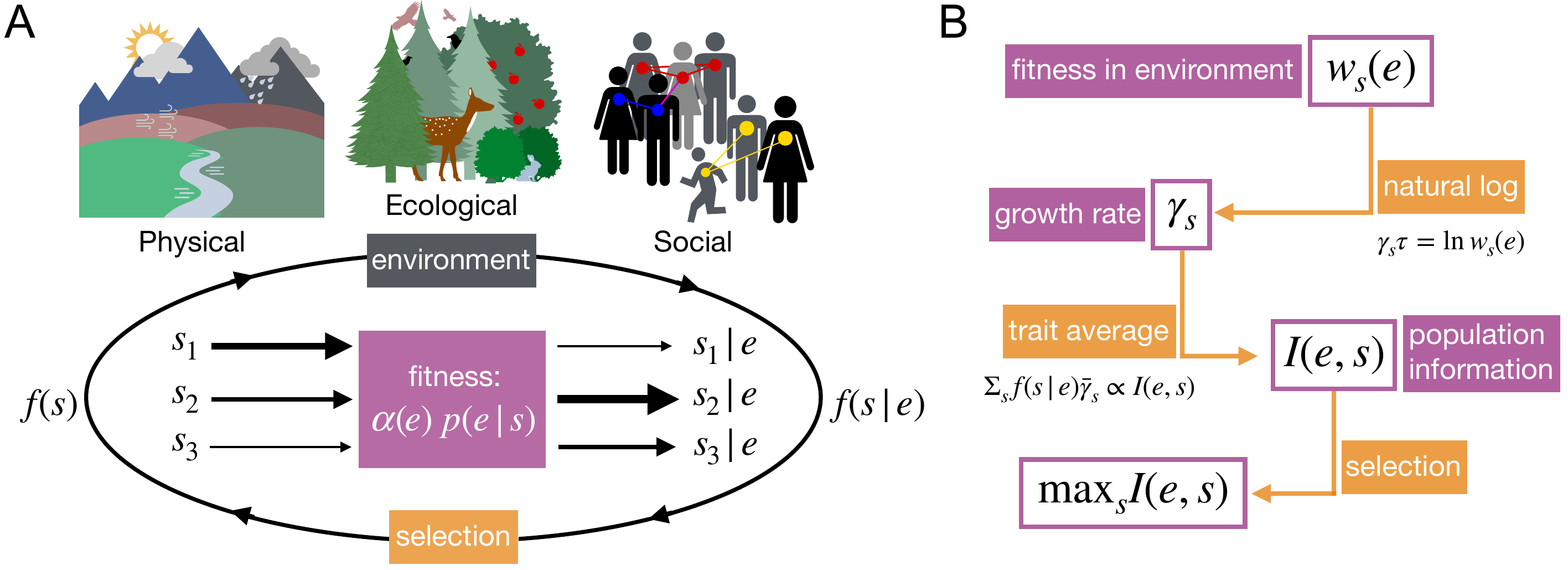}
  \caption{\textbf{Scheme of evolutionary population dynamics of selection in various environments. }\textbf{A}: The fitness as a model of adaptation becomes proportional to a likelihood of environmental states given the phenotypes. \textbf{B}: The iterated process produces time averaging over the environment statistics and selects the population types that maximize the average log-fitness, which converges to the mutual information between the population structure and the environment.}
  \label{fig:Fig-1}
\end{figure*}

\subsection*{Selection, Bayesian learning and fitness}
Fitness refers to (optimal) adaptation to an external environment, $e$. However, Eq.~\ref{eq:fitness} and the standard set-up of population genetics~\cite{crow1970introduction} make no explicit reference to $e$: We now show how these perspectives can be reconciled.

In evolutionary dynamics, the focus of optimization is the maximization of relative fitness. The time evolution of type frequencies follows from Eq.~\ref{eq:fitness} as
\begin{eqnarray}
  f_s(t)=\frac{N_s(t)}{N(t)}= \frac{w_s }{w} \frac{N_s (t-1)}{N (t-1)}=\frac{w_s }{w} f_s(t-1). 
  \label{eq:frequency-update}
\end{eqnarray}
The mapping to Bayes' theorem follows immediately as 
\begin{eqnarray}
    f_s(t)=\frac{w_s }{w}f_s(t-1)  \rightarrow 
  p(H|D) =\frac{p(D|H)}{p(D)} p(H).
  \label{eq:frequency-to-Bayes}
\end{eqnarray}

Several authors have recently noted that the replicator equation under pure selection, Eq.~\ref{eq:frequency-update}, maps exactly to Bayes relation~\cite{harper2009replicator,akyildiz_replicator_2017,Czegel2022, czegel_multilevel_2019}, written on the right side of Eq.~\ref{eq:frequency-to-Bayes}. Bayes theorem gives the update of the probability that a hypothesis $h \in H$ is true, given an initial distribution $p(h)$ (the prior), and new evidence $D$. This is written as the updated (posterior) probability $p(H|D)$ over all $h \in H$. Because Bayes' theorem is an  identity that follows from the rules of probability, it is exact and the optimal way to incorporate new evidence~\cite{behrens2007learning}. As $p(H|D)$ is more constrained than $p(H)$ the data (environment) effectively selects specific hypotheses as more likely in a process where information is learned from the data about the hypothesis space. 

The key quantity in this procedure is $p(D|H)$, known as the likelihood. Although it looks like another probability, the likelihood evaluated over the data and the hypotheses is the engine of selection. It is interpreted in Bayesian estimation as a probabilistic model predicting the data given each hypothesis. The remaining factor is a normalization because $p(D)=\sum_h p(D|h)p(h)$. 

Thus, the interpretation of type frequencies as probabilities leads to the mapping of evolutionary dynamics under selection to Bayesian learning~\cite{harper2009replicator,Czegel2022,czegel_multilevel_2019}. This is a condition of adaptive optimality, in the sense that it gives the most predictive change in the relative population size of various types, given events in the environment~\cite{frank2009natural}. 

This identification suggests a redefinition of fitness as an explicit probabilistic model of environmental adaptation without additional assumptions. To see this, let us map the data, $D$, to environmental states, $D\rightarrow e$, and hypotheses to population types, $H \rightarrow s$. We then identify $f_s(t) = f(s|e)$: at each time step $t$, the environment is observed in state $e$, so the post-selection frequency of type $s$ equals the Bayesian posterior probability of type $s$ given that environmental observation. This leads to
\begin{eqnarray}
  f(s|e)= \frac{p(e|s)}{{\hat p}(e)} f(s) \rightarrow w_{s}(e) = \alpha(e) p(e|s), \quad 
\end{eqnarray}
where the fitness can be expressed in terms of the likelihood, and the average fitness becomes
\begin{eqnarray}
 w(e)=\sum_s w_{s}(e) f(s) = \alpha(e) \sum_s p(e|s) f(s)=\alpha(e){\hat p}(e).  \label{eq:fitness-as-likelihood} 
\end{eqnarray}

This leads to the identification of fitness with the Bayesian likelihood, $w_{s}(e)=\alpha(e) p(e|s)$. The factor $\alpha(e)$ is an environment-dependent normalization ensuring that the likelihood $p(e|s)\leq 1$ maps to a valid absolute fitness. To see this, rewrite eq. \ref{eq:fitness} as
 \begin{eqnarray}
N(t)=\sum_s N_s (t) = \sum_s w_{s}(e) N_s(t-1) =  \sum_s  \alpha(e) p(e|s) N_s (t-1) = \alpha(e){\hat p}(e) N(t-1),
\label{eq:fitness2}
\end{eqnarray}  
which shows that $\alpha(e)>1/p(e|s)$ is necessary for $N_s$ to grow over time, and similarly $\alpha(e)>  1/{{\hat p}(e)}$ for the total population.  The factor of $\alpha(e)$ disappears, however, in relative fitness and does not affect (relative) selection. Note that ${\hat p}(e)$  is the estimate of the true environmental probability, $p(e)$, obtained by averaging the likelihoods across types. 

We have now achieved a definition of fitness for each type $s$ as an explicit probabilistic model of the environment. This definition resolves the three conceptual problems raised in the Introduction.  Circularity is broken because fitness is now a statistical model of a specific (statistical) environment. Different environments will be better predicted by different such models belonging to different types, resulting in selection. The model is predictive in the sense that phenotypes anticipate their environmental states. The Bayes classifier codifies optimality because it is the best possible predictor of the environment~\cite{mackay_information_2003}, see below.  The mismatch problem thus vanishes because $\alpha(e) p(e|s)$ reflects probabilistic adaptive capacity independently of any particular event. Importantly, this also implies that our definition of fitness will require averaging over events in the environment.  The propensity interpretation of fitness is naturally operationalized because $p(e|s)$ is inherently probabilistic and predictive.  The picture of adaptation that emerges is explicitly relational and statistical: selection relies on diverse phenotypes as alternative generative models of the environment, $p(e|s)$, whose fidelity to relevant environmental statistics drives relative reproductive success.

\subsection*{Fitness, growth rates, and information}
The fitness defined in Eq.~\ref{eq:fitness-as-likelihood} is a stochastic quantity because it is evaluated at actual states of the environment. Consequently, we must discuss how to average it and what quantities are maximized by the population dynamics of Eq.~\ref{eq:frequency-update}. The connection between fitness, probability and information comes into focus when we consider that
\begin{eqnarray}
  \ln w_{s}(e) = \ln \alpha(e) + \ln p(e) + \ln \frac{p(e|s)}{p(e)}, 
\label{eq:information}
\end{eqnarray}
which is the instantaneous growth rate, $\gamma_s$. 
The first term shows the payoffs from the environment, the second measures environmental variability (or surprise~\cite{mackay_information_2003}), the third term measures how close the $s$-type model of the environment is to the real thing, $p(e)$, as we show next. Only this last term depends on the type and contributes to relative fitness. 

When the population dynamics is iterated over a (long) time $T$, the types with the highest average log-fitness over such period are positively amplified. This emerges from the temporal dynamics over $T$ because
\begin{eqnarray}
w_{s} (E) = \prod_{t=1}^T A(e_t) p(e_{t}|s)  \rightarrow \ {\bar \gamma}_s = \frac{1}{T} \sum_{t=1}^T \ln \!\left[ A(e_t) p(e_{t}|s) \right].
\end{eqnarray}
It follows that
\begin{eqnarray}
    {\bar \gamma}_s  \simeq \ln \alpha + \sum_e p(e) \ln p(e) + \sum_e p(e) \ln \frac{p(e|s)}{p(e)},
\end{eqnarray}
or equivalently in terms of information quantities,
\begin{eqnarray}
    {\bar \gamma}_s = \ln \alpha - H(E) - D\!\left(p(E)\,\|\,p(E|s)\right), \label{eq:growth-info}
\end{eqnarray}
where $H(E)=-\sum_e p(e)\ln p(e)\geq 0$ is the Shannon entropy of the environment and $D(p(E)\|p(E|s))\geq 0$ is the Kullback-Leibler (KL) divergence~\cite{hledik_accumulation_2022}. Here, ${\bar \gamma}_s$ is the average growth rate of type $s$ over a temporal history (lineage); the approximation converges to the actual rate when the long-term temporal trajectory averages over $e$ by the law of large numbers. The quantity $\ln \alpha$ is the average of $\ln \alpha(e)$ (see Supplementary Information), a property of the environment setting the magnitude of the largest possible growth rate. The entropy $H(E)$ is always positive and larger in more complex environments, which therefore entail a growth penalty because they are harder to predict. The KL divergence is non-negative and equals zero only when $p(e|s)=p(e)$, so selection amplifies the type $s$ with the most accurate probabilistic model of the environment; all other types incur a relatively smaller growth rate. Thus, the population dynamics selects the types $s$ that maximize the average growth rate~\cite{taylor2007information,kemp2023learning}. Averaging type growth rates over the posterior $f(s|e)$ yields the mutual information $I(E,S)$ between population structure and environment (see Supplementary Information for derivation):
\begin{eqnarray}
\sum_{s,e} f(s|e)\,p(e)\,\ln w_s = \ln \alpha - H(E) + I(E,S). \label{eq:mutual-info}
\end{eqnarray}
In this sense, selection (Eqs~\ref{eq:fitness}--\ref{eq:fitness-as-likelihood}) converges on the type distribution that maximizes the mutual information $I(E,S)$ between population structure across types and the statistics of the environment.

Two additional consequences follow from the definition of fitness as a Bayesian likelihood. First, because fitness is (proportional to) a probability, statistically independent contributions combine as products. This means that fitness will only rarely be a linear combination of factors, as often assumed. Second, statistically independent contributions do add up in the log-fitness, as in all information quantities~\cite{mackay_information_2003}. This makes the average log-fitness (or growth rate) the natural quantity for analyzing complex evolutionary problems, especially away from weak selection.  Next we show that analysis of the growth rate also predicts phase transitions in evolution, with environmental averages playing the role of control parameters.

\subsection*{Fitness in probabilistic environments} 
We now illustrate how the Bayesian likelihood framework creates analytical tools for three canonical problems: task switching (how selection tracks environmental statistics), evolutionary games (how log-fitness optimization yields Hamilton-type cooperation conditions), and selection in groups (how group structure modifies these conditions through in-group and out-group interactions).

\subsubsection*{\bf Task switching} 
In the problem of task switching, a population of agents is distributed to a set of discrete tasks, $s$, each associated with differential payoffs dependent on environmental states, $e$. Optimal evolutionary dynamics align these tasks with the environment to maximize rewards. For simplicity, we consider two tasks, denoted $s=0,1$, in a binary environment, $e=0,1$. The standard setup prescribes rewards, $Y_{s}(e)$, to each state combination, Table~\ref{tab:Table_1}A-B. We show that the long-run selected population distribution mirrors the environmental probability, $f(s)\to p(e)$, maximizing $I(S,E)$.

 To understand the dynamics of selection, note that it pays off to align the task with the environment $s=e$, regardless of the specific state, $e=0,1$ because $Y_{s=e}=Y_+> Y_{s\neq e}=Y_-$. The question becomes interesting when the environment is only known in probability, $p(e)$, assumed stationary. Then, starting with some probability of tasks $f(s)$, the population is selected over time, updating the conditional probability $f(s|e)$ as instances of the environment are experienced. 

The identification of population dynamics of this type with iterated Bayesian selection requires mapping the rewards $Y_{s}(e)$ to the likelihood function, $p(e|s)$. Because the likelihood is a conditional probability, it is positive and normalized, meaning that any mapping to payoffs must satisfy such requirements. We want mappings that preserve the {\it relative} strength of the payoffs. 
This is achieved by writing the fitness, $w_{s}(e)=\alpha p(e|s) = Y_{s}(e)$. This requires that $Y_{s}(e) > 0$ as assumed in Table~\ref{tab:Table_1} and that the factor of $\alpha(e)$ is defined to ensure that none of the probabilities exceeds one and that the conditional probability is normalized. In general, the factor of $\alpha$ is a function of the state of $e$, but since the example of Table~\ref{tab:Table_1} is symmetric, $\alpha=5/2$ and $
p(e=0|s=e) = p(e=1|s=e) = \frac{4}{5}$ and $ p(e=0|s\neq e) = p(e=1|s\neq 0) = \frac{1}{5}$. We can write the fitness as 
\begin{eqnarray}
w_s(e) = Y_+^{(1-s)(1-e)+se} Y_-^{(1-s)e + s(1-e)}.
\end{eqnarray}
The logarithm of the fitness (log-fitness) is then
\begin{eqnarray}
\ln w_s = \ln Y_+ + \ln \frac{Y_+}{Y_-} \left( 2 se - s - e \right).
\end{eqnarray}
We can now obtain the probability of $s$ that maximizes fitness over a long (stationary) history of the environment by taking expected values over $e$ and $s$. We define $f(s=1)=p$ and $p(e=1)=q$, and $f_{1|1}=f(s=1|e=1)$ and $f_{1|0}=f(s=1|e=0)$. The log-fitness is maximized when the quantity $(2 f_{1|1} -1)q - p$ is maximal. This requires $f_{1|1}=1$, which implies $p=f_{1|1}q+f_{1|0}(1-q)=q$. Hence, the population task distribution evolves to mirror the environmental probability. 

We can also write the dynamics of selection by iterating the likelihood, given an environmental history and an initial distribution of traits $s$, see Supplementary Information
\begin{eqnarray}
f(s|e) \sim \Pi_{t=1}^T p(e_t|s) f(s) = Y_+^{N_{e=s}}Y_-^{T- N_{e = s}} f(s),
\end{eqnarray}
where $N_{e=s}$ is the number of instances that $s=e$ over $T$ times. For long times, the law of large numbers gives $N_{e=s}/T \rightarrow p(e=s)$, the probability that the environment matches the type. The likelihood is maximized by taking $p(s=e|e) \rightarrow 1$, which means that $f(s) \rightarrow p(e)$, as derived above. 

We conclude that the state of the population that maximizes the (Bayesian) dynamics of selection exactly mirrors that of the environment. The state of $s$ selected at long times also maximizes the information that the population has on the environment. This follows from $I(S,E) = H(S) - H(S|E)=H(E)$, meaning that the variation of the environment is now fully explained by population statistics. In this sense, evolutionary dynamics selects the state that maximizes information transmission between the environment and the population.

\subsubsection*{Evolutionary games}
Evolutionary game theory (EGT) provides a general framework for the study of coevolutionary dynamics resulting from social interactions. The population now becomes an (adaptive) environment from which each agent derives its fitness. This contrasts with the task selection problem, where environmental dynamics are stationary and therefore independent of the agent's actions.  

We first demonstrate how the statistical treatment of fitness simplifies the analysis of dyadic games. Consider the prisoners' dilemma (PD), used for decades to analyze the evolution of cooperation~\cite{Trivers1971,Axelrod1980}. Table~\ref{tab:Table_1}C-D shows the game's payoff structure, where $s=0$ denotes defection and $s=1$ cooperation. We consider two players $s_1, s_2$ each with cooperation probability $p_1, p_2$, respectively. Payoffs are such that $W>R>L>P>0$. We write $s_1$'s fitness as
\begin{equation}
w_{s_1} (s_2)= \alpha R^{s_1 s_2} W^{(1-s_1)s_2} P^{s_1(1-s_2)} L^{(1-s_1)(1-s_2)}.
\label{eq:fitness-games}
\end{equation}
Note that this is different from the standard EGT replicator equation~\cite{gintis2000game}.  Eq.~\ref{eq:fitness-games} gives stochastic pay-offs at each interaction rather than on average over the population~\cite{gintis2000game,hofbauer_evolutionary_1998}.  In this sense, we start from not assuming a well-mixed population: the familiar replicator equation with linear, frequency-weighted payoffs is recovered only as an infinite-population, time-averaged (mean-field) limit of Eq.~\ref{eq:fitness-games}, and corresponds to the statistically independent ($r=0$) case of Table~\ref{tab:Table_1}E (see SI, \emph{The standard EGT replicator equation is a mean-field approximation}).  This leads to 
\begin{equation}
\ln w_{s_1} = a - c s_1 +b s_2 + d s_1 s_2,  
\label{log-fitness_PD}  
\end{equation}
where $a=\ln \alpha L$, $c= \ln \frac{L}{P}$, $b=\ln \frac{W}{L}$, and $d= \ln \frac{R L}{P W}$. This polynomial log-fitness coincides with the typical parameterization of fitness as a linear function of costs and benefits of cooperation in the classical problem of altruism~\cite{Trivers1971,Axelrod1980}. Specifically, there is a fitness cost, $c>0$, of cooperating ($s_1=1$), and an independent benefit $b>0$ from the cooperation of others ($s_2=1$). There is also an extra benefit (if $d>0$) when both agents cooperate, which was considered by Queller~\cite{Queller1985,Queller1992}. Note also that $d= c -\delta <c$, with $\delta = \ln(W/R)>0$. In a single-shot game, $s_1=0$ (defection) results in higher log-fitness even when the other player cooperates ($s_2=1$). This is the game's familiar Nash equilibrium, because the log-fitness of $s_2$ is identical with labels interchanged.

\begin{table}[ht]
\centering
\small
\begin{tabularx}{\textwidth}{X X} 
  \vspace{.5em}
  \textbf{A}
  \vspace{-2em}
  \begin{center}
  \begin{tabular}{c c c}
    \hline
     Fitness, $w_s(e)$ & $e=0$ & $e=1$ \\
    \hline
    $s=0$ & $\alpha p(0|0)$ & $\alpha p(1|0)$ \\
    $s=1$ & $\alpha p(0|1)$ & $\alpha p(1|1)$ \\
    \hline
  \end{tabular}
  \end{center}

  \textbf{B} 
  \vspace{-1.75em}
  \begin{center}
  \begin{tabular}{c c c}
    \hline
    Returns, $Y_{s}(e)$ & $e=0$ & $e=1$ \\
    \hline
    $s=0$ & $Y_+$ & $Y_-$ \\
    $s=1$ & $Y_-$ & $Y_+$ \\
    \hline
  \end{tabular}
  \end{center}

 &
  
  \vspace{.5em}
\textbf{C} 
  \vspace{-2em}
  \begin{center}
  \begin{tabular}{c c c}
    \hline
    $Y_{s_1,s_2, e=0}$ & $s_2=0$ & $s_2=1$ \\
    \hline
    $s_1=0$ & $L,L$ & $W,P$ \\
    $s_1=1$ & $P,W$ & $R,R$ \\
    \hline
  \end{tabular}
  \end{center}
  
\textbf{D}
  \vspace{-1.75em}
  \begin{center}
  \begin{tabular}{c c c}
    \hline
    $Y_{s_1,s_2, e=1}$ & $s_2=0$ & $s_2=1$ \\
    \hline
    $s_1=0$ & $L,L$ & $P,W$ \\
    $s_1=1$ & $W,P$ & $R,R$ \\
    \hline
  \end{tabular}
  \end{center} \\

\vspace{.25em}
\textbf{E} \hspace{1.5em}
  \begin{tabular}{p{0.25\textwidth} p{0.3\textwidth} p{0.3\textwidth}}
    \toprule
    \textbf{Relation} & \textbf{Cooperation Condition} & \textbf{Parameters} \\
    \midrule
    Statistical independence (SI) & never &  \\
    Well-mixed population (WM)   &  $r_{\rm m}~b+f_{\rm m}~d > c$  & $r_{\rm m}=1$, $f_{\rm m}=p$  \\
    Conditional dependence (CD)  & $r_{\rm d}~b + f_ {\rm d}~d >c$ & $r_{\rm d}=f_{1|1} - f_{1|0}$, $f_{\rm d}= f_{1|1}$ \\
    Mixed games & $r_{\rm d} ~b(q) + f_{\rm d} ~d > c(q)$ & $q=p(e=1)$  \\
    Selection in groups    &  $ \frac{k_g} {K} \left( r_g ~b + f_g ~d \right) + d p_{\bar g} >  c $ &  $r_g= f_{1g|1g} - f_{1g|0g}$, $f_g= f_{1g|1g} - p_{\bar g} $ \\
    \bottomrule
  \end{tabular}
\end{tabularx}

\caption{Combined payoff tables for the task switching problem and 2-agent games. 
\textbf{A}: \textnormal{Payoffs for task alignment with the environment, where $Y_+=2,Y_-=1/2$. }
\textbf{B}: 
\textnormal{Fitness mapping to likelihood. }
\textbf{C}:
\textnormal{Payoffs for the prisoners' dilemma ($e=0$) with $W>R>L>P$. }
\textbf{D}:
\textnormal{Payoffs for the cooperation game ($e=1$) with $W'=P, \ P'=W$, inverting the prisoner's dilemma cross strategy payoffs. }
\textbf{E}: 
\textnormal{Summary of conditions for the emergence of cooperation.}}
\label{tab:Table_1} 
\end{table} 

By contrast, the goal of evolutionary game theory is the identification of selected strategies over long histories, which is analyzed by averaging Eq.~\ref{log-fitness_PD}. The key variable is the statistical relationship between the two agents' strategies. We examine three cases of increasing complexity — statistical independence, a well-mixed population (agents share the same marginal cooperation probability), and conditional dependence (agents' strategies are correlated) — each yielding a qualitatively different outcome.

When the agents are statistically independent with cooperation probabilities $p_1, p_2$, the average log-fitness is
\begin{eqnarray}
\ln w_{s_1} = a - c p_1 + b p_2 + d p_1 p_2 = a - \left[ c (1-p_2) + \delta p_2 \right] p_1 + b p_2.
\label{eq:basic}
\end{eqnarray}
The quantity in the square bracket is always positive, so $p_1=0$ maximizes the fitness independently of $p_2$. This means that the agent should always defect ($s_1=0$), as expected. Here, statistical independence means intuitively that the other agent does not react, even when it is exploited. An example is the single shot game, which has the same optimal solution. Note also that statistical independence means that there cannot be any information between the agents. 

\begin{figure*}
  \centering
  \includegraphics[width=\linewidth,scale=0.6]{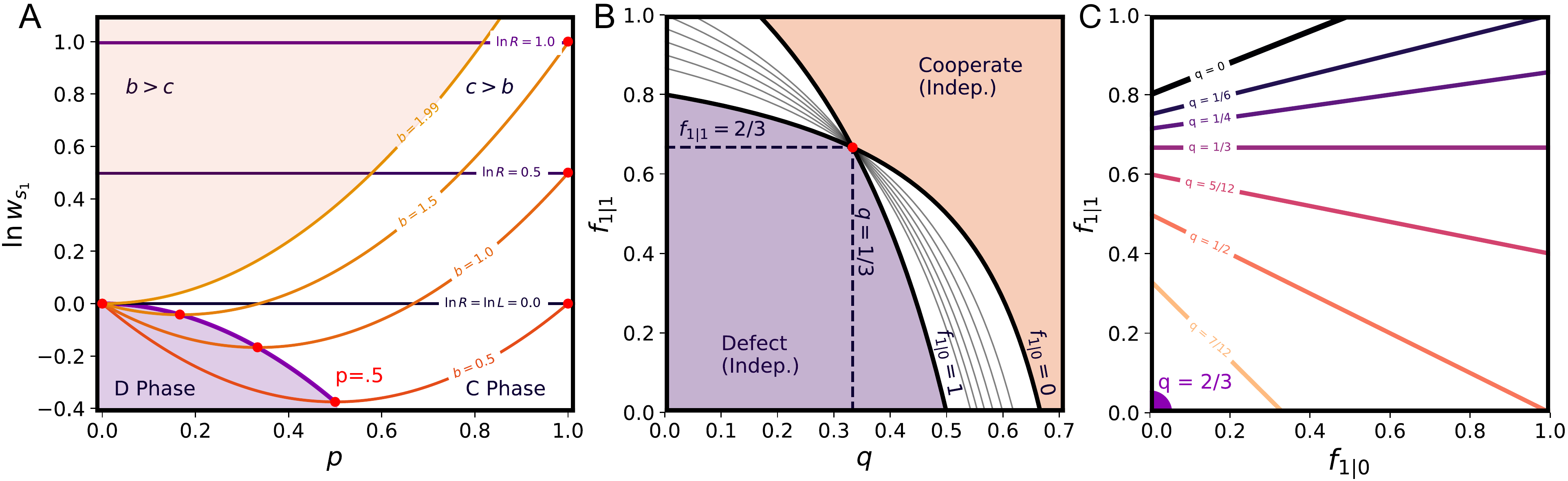}
  \caption{\textbf{ Phase diagram for the Prisoner's Dilemma and the environmentally mixed game.} \textbf{A}: Log-fitness versus cooperation probability $p$ for the well-mixed population. Note the fitness valley for $b<c$. Conditional dependent case conditions for cooperation. \textbf{B}: Probability of reciprocity $f_{1|1}$ versus environment, $q$. \textbf{C}: Probability of reciprocity versus exploitation $f_{1|0}$, for various $q$. Cooperation emerges in each case for parameters above the lines. See also~Fig. 1 of the supplement}
  \label{fig:Fig-2}
\end{figure*}

Now consider a seemingly small but important change in these assumptions. We consider $p_1=p_2=p$, which is frequently assumed in population genetics. This is often dubbed random mating (in diploid populations) or a fully mixing population~\cite{crow1970introduction}. This case --- in which the two agents share the same marginal, so that the population effectively plays itself ($r_{\rm m}=1$ in Table~\ref{tab:Table_1}E) --- is distinct from, and more favorable to cooperation than, the random-matching assumption (statistical-independence, $r=0$) that underlies the standard replicator equation (SI). It results in
\begin{eqnarray}
\ln w_{s_1} = a - (c-b) p + d p^2. 
\label{eq-random-mixing}
\end{eqnarray}
The quadratic form of the log-fitness in $p$ is familiar as an effective potential, and simple to analyze. A distinction emerges depending on the relative magnitude of $b$ and $c$, Fig.~\ref{fig:Fig-2}A. When $b>c$ (or $WP > L^2$) the situation is simple because the log-fitness is always a growing function of $p$, with a single maximum at $p=1$ corresponding to full cooperation. Thus, whatever the initial state of the population, cooperation will always emerge as the dominant strategy and will eventually fixate. Relative to the statistically independent case, cooperation now emerges from the fact that the population is effectively playing itself, so that increases in mutual exploitation lead to overall lower fitness. The situation is more interesting when $c>b$, because the average log-fitness has a minimum at $p_{\min} = \frac{c-b}{2d}$. For $p < p_{\min}$ the fitness decreases with $p$, so mutual defection ($p=0$) is locally stable; for $p > p_{\min}$ the fitness increases toward $p=1$. Thus $p_{\min}$ is the dynamical tipping point: sufficient initial cooperation $p > p_{\min}$ is needed for the population to converge to full cooperation ($p=1$), Figure~\ref{fig:Fig-2}A. Full cooperation is the global maximum when $\bar\gamma(1) > \bar\gamma(0)$, which requires $d > c-b$. Defining $p_* \equiv \frac{c-b}{d}$, this condition is simply $p_* < 1$. From Eq.~\ref{eq-random-mixing} we identify total benefits at full cooperation $B = b+d$ and costs $C=c$; then $B>C$ is the condition for cooperation to be globally optimal, which is a form of Hamilton's rule~\cite{Hamilton1964}. The population dynamics around $p_{\min}$ is characterized by a first-order transition, a "tipping point," because starting with small $p$ there is a valley of fitness to be crossed to reach the global maximum.  This can be done locally, over space or on networks, as many authors have noted based on numerical experiments~\cite{nowak2004emergence,allen2014games,chen2015competition}. We will return to this point below. 

Third, the situation is more interesting and more complex when the agents' actions become statistically dependent. This can arise in a variety of ways, such as in repeated games or in the presence of signaling.  In this case, there are three parameters at play in the joint distribution: four joint probabilities minus the normalization. We can parameterize this space in analogy to task selection as $p_2 = f_{1|0} (1-p_1) + f_{1|1}p_1$ and $f(s_1=1,s_2=1)=f_{1|1} p_1$, with the shorthand $f_{1|1}= f(s_2=1|s_1=1)$, $f_{1|0}= f(s_2=1|s_1=0)$. Note that $r=f_{1|1}-f_{1|0}$ is the regression~\cite{frank_natural_2012,Queller1992} of $p_2$ on $p_1$, which vanishes when the two agents are statistically independent. Though this is a general statistical quantity, it arises naturally from kinship~\cite{Hamilton1964}.   Thus, $r$ is a measure of how (co-)related the two agents' phenotypes are. The log-fitness becomes
\begin{eqnarray} 
\ln w_{s_1} = a + b f_{1|0} + [  r b +d f_{1|1} - c ] p_1.
\label{eq-cond-polynomial}
\end{eqnarray}
Because this is linear in $p_1$ the nature of the maximum depends only on the sign of the term in the square brackets. This is a second version of Hamilton's rule closer to the way Queller~\cite{Queller1992} expressed it. 
If we identify the benefits of interaction $B=b$, and costs $C=c$, we now obtain $r B>C$ as the condition for cooperation to emerge. Adding the $d$ term facilitates the emergence of cooperation via what Queller called a synergism coefficient, as $rB + d f_{1|1} >C$.  This transition has a geometric interpretation in the reciprocity-exploitation ($f_{1|1}, f_{1|0}$) plane. We can write the cooperation condition as a straight line, $y=f_{1|1}$, $x=f_{1|0}$, $ y = y_0 + \beta x$, with $y_0 = \frac{c}{b+d}$, $\beta=\frac{b}{b+d}$ , as shown in Figure~\ref{fig:Fig-2}C. Cooperation, $p_1\rightarrow 1$, emerges for values above this line. To be above the line requires controlling exploitation, optimized when $f_{1|0}=f(s_2=1|s_1=0) \rightarrow 0$, which implies that punishment or retaliation is strong, $f(s_2=0|s_1=0) \rightarrow 1$. It also requires reciprocity, meaning that $f_{1|1} = f(s_2=1|s_1=1) \rightarrow 1$, so that one becomes safe from exploitation, $f(s_2=0|s_1=1) \rightarrow 0$. Note that the first of these conditions may take the population towards mutual defection~\cite{rand2010anti}, which is suboptimal. The second is therefore necessary to reach the highest fitness globally. Below the line, cooperation cannot emerge. These outcomes agree with established results on the effects of reciprocity, punishment and rewards on cooperation~\cite{sigmund2001reward, hintze2010darwinian}. That is, both punishment and reciprocity promote the establishment of cooperation but alone punishment does not suffice~\cite{nakamaru2009runaway, rand2010anti}. The information in the population is maximal as agents coordinate but the amount of information at stake converges to zero at full cooperation because then there is no uncertainty left to explain. 

All other 2x2 games can be analyzed in the same way, see Supplementary Information. Phase transitions in population dynamics are expected in general as the polynomial terms in the log-fitness change sign, driven by external parameters such as characteristics of the environment, as we show next. 

\subsubsection*{ \bf Mixed games in variable environments}
 Greater realism and more possibilities result from combining several interaction strategies presented by the environment. Here, we illustrate this situation via the mixture of two games, PD for $e=0$, and "turn the other cheek" (ToC), for $e=1$. In ToC, defecting against a cooperator is costly and the highest payoff comes from cooperating with a defector — it is a game of conspicuous virtue, obtained by inverting PD's off-diagonal payoffs, Table~\ref{tab:Table_1}C-D.

Selected strategies depend on the relative probabilities of the two games set by the environment $q=p(e=1)$, which is the probability of ToC.  The log-fitness becomes
\begin{eqnarray}
\ln w_{s_1} = a - c(e) s_1 + b(e) s_2 + d s_1 s_2,
\label{eq:mixed-game}
\end{eqnarray}
with $c(e)= c (1-e) - b e$ , $b(e) = b(1-e) - c e$. Costs and benefits swap between the PD and the ToC, for which $\ln w_{s_1} = a + b s_1 -c s_2 + d s_1 s_2$. If the two agents act independently, then the agent should always cooperate in the ToC because both $b, d > 0$. If the two agents share the same probability $p_1=p_2=p$, the log-fitness is independent of $e$ because there is a symmetry between the two agents so that the benefits and costs of cooperation are fully socialized. Thus, in the fully mixed population, changes in the environment have no effect on the average dynamics and the phase diagram is the same as in Fig.\ref{fig:Fig-2}A.

When the two agents coordinate, the situation becomes more interesting. If agents can tell which game they are playing at each time, their probability of cooperation will simply follow from the selected strategy for each game, weighed by their relative probability, $q$. Alternatively, we consider when agents act conditionally on each other, independently of the state of the environment. Then we obtain the same condition for cooperation to emerge as in PD, but with costs and benefits that depend on the environment as in Eq.~\ref{eq:mixed-game}. Fig.~\ref{fig:Fig-2}B-C shows that when $q=0$, we obtain the PD condition (black line) in Fig.~\ref{fig:Fig-2}. As $q$ increases, both the ordinate at the origin and the slope decrease. In general, the slope vanishes at $q=b/(b+c)$ and cooperation becomes unconditional for $q \geq c/(b+c)$ (e.g., $q=1/2$ and $q=2/3$ for the parameters shown).

This analysis is general and can be obtained for any combination of two-player games, leading to complex and interesting phase spaces, Fig.~\ref{fig:Fig-2}. As a result, we obtain transitions between different (optimal) population states as functions of environment probabilities, which play the role of control parameters analogous to temperature or pressure in physical systems. We have therefore shown how mixing the prisoner's dilemma with a cooperative game increases the selection for cooperation in non-trivial ways. This is likely a more realistic situation as organisms, and especially people, have a diversity of modes of interaction, some competitive and some cooperative. Thus, the presence of explicitly cooperative channels some of the time helps unstable cooperation to emerge all of the time. 

\subsubsection*{\bf Selection in groups}
The Bayesian fitness framework extends naturally to structured populations. We show that the Hamilton-type cooperation conditions derived above generalize to this setting, with in-group and out-group interactions entering as distinct control parameters. Group identities specify additional layers of organization that condition phenotypes and therefore contribute to individual selection.

We write the fitness of individual $i$ in group $g$ as $ w_{gi} (g',j)= \alpha p(\{ s_{g' j} \} |s_{g i})$,
where $\{ s_{g' j } \}$ denotes the joint state of all neighbors $j, g'$. Typically, these neighbors are not statistically independent, leading to synergistic interactions~\cite{Queller1985,bettencourt2009rules}. This may be important for finding clusters of states (genes, phenotypes) that collectively encapsulate adaptive functions~\cite{kemp2024information}. 

However, the most common situation arises when interactions are dyadic (between pairs only), so that the game occurs on a simple graph \cite{Ohtsuki2006,e_james_lieberman_evolutionary_2005,szabo2007evolutionary,allen2014games}. 
In the log-fitness, this approximation is obtained when all the $s_{g' j}$ are conditionally independent given $s_{g i}$ such that $w_{gi} = \Pi_{j,g'} \alpha p( s_{g'j} |s_{gi})$, resulting in
\begin{eqnarray}
\ln w_{g i} = \ln \alpha + \sum_{g',j} \ln p( s_{g'j} |s_{gi}).
\end{eqnarray}
Under these circumstances, we can write the log-fitness to be identical to 
Eq.~\ref{eq:basic}, with the second player replaced by the average over all connections, $s_2=\frac{1} {K} \sum_{g' ,j} {s_{g' j}}$. All our previous results still apply but require attention to group dependencies, see Supplementary Information for details. 

\begin{figure*}
  \includegraphics[width=\linewidth]{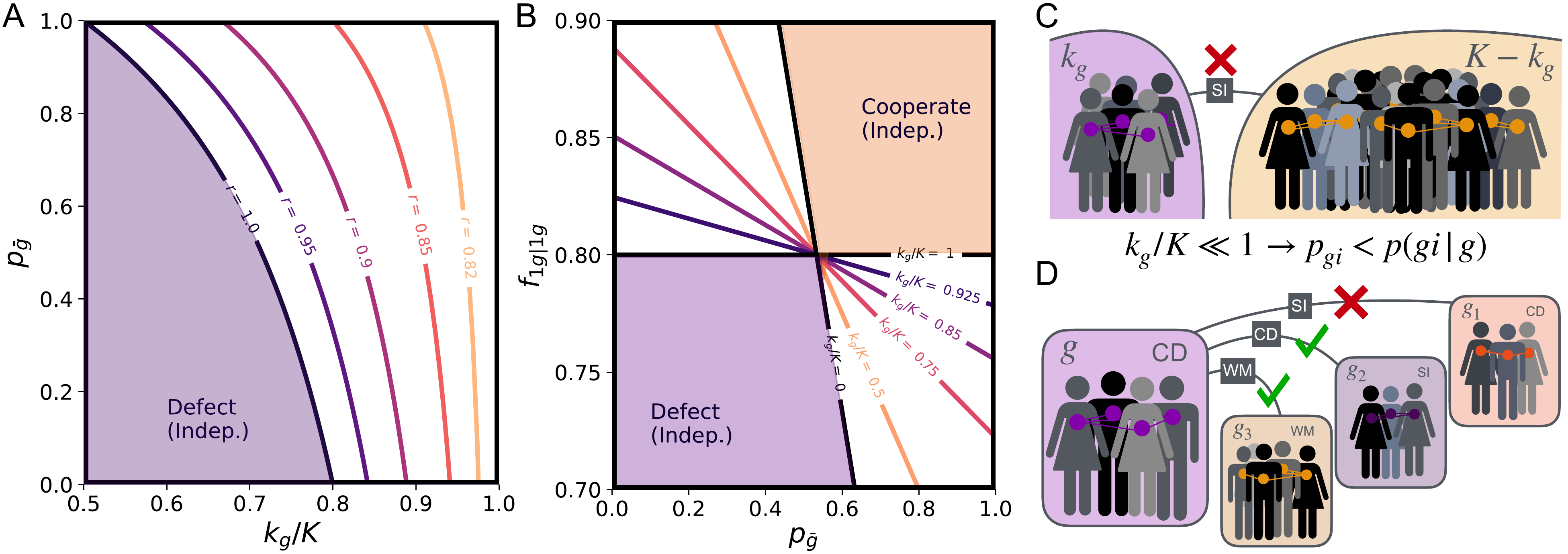} 
  \captionof{figure}{\textbf{Selection in structured populations depends on group sizes and treatment of out-groups. }\textbf{A}: Phase diagram as a function of cooperation from outer groups $p_{\bar g}$ and fraction of connections within in-group, $k_g/K$, for various amount of in-group relatedness $r$. \textbf{B}: Phase diagram as a function of cooperation probability in out-groups and in group reciprocity, for different fractions of in-group connectivity, $k=k_g/K$.  
  Cooperation emerges for conditions above each line.
  \textbf{C}: Treating out-groups as independent promotes defection across groups and in the general population. \textbf{D}: Cooperation can be expanded as interactions become cooperative across group labels.} 
  \label{fig:Fig-3}
\end{figure*}
 
In particular, the agent should still defect unconditionally when it is independent from the other agents. When their probabilities coincide, in the well-mixed population $p_{gi}=p_2 = p$, Eq.~\ref{eq-random-mixing} still applies and a phase transition occurs at $p=p_{\min}=(c-b)/(2d)$, if $b<c$. When the states are not independent, the log-fitness is written compactly in terms of contributions from each interaction, weighed by population size $1/K$, 
\begin{eqnarray}
\frac{1} {K} \ln w_{g i} = a + b f_{1|0g} + [ r b +d f_{1|1g}  - c ] p_{gi}.
\end{eqnarray}
This equation is similar to what we obtained above, but with population definitions of $r$, $f_{1|1g}, f_{1|0g}$, where the additional $g', g$ labels denote conditional dependence across groups; the absence of the label means average across groups. The interesting possibility is that the agents' behaviors are now different within and across groups in the sense that reciprocity $f_{1g'|1g} < f_{1g|1g}$, and exploitation
$f_{1g'|0g} > f_{1g|0g}$, for $g' \neq g$. Treating out group interactions as independent leads to 
\begin{eqnarray}
\frac{1} {K} \ln w_{g i} =& a + \underbrace{ \frac{k_g}{K} \left( b r_g + d f_{1g|1g} -c  \right)}_{\rm in-group} p_{gi} + \underbrace{ \sum_ {g'\neq g} \frac{k_{g'}}{K}\big[ \left( d p_{g'} -c  \right)p_{gi} +b
p_{g'} \big]}_{\rm out-groups} .
\end{eqnarray} 
This presents an interesting conundrum: The in-group contribution drives cooperation ($p_{g i} \rightarrow 1$), but the out-group selects for defection($p_{g i} \rightarrow 0$). In general, the out-group is larger so cooperation may be corrupted by the temptation to exploit out-group members, or simply by the failure to coordinate phenotypes.

This collapse of cooperation under external pressure (or temptation) can be averted in two ways. First, the agent may cease to interact with out-group members so that $k_{g'\neq g}=0$, but this foregoes fitness benefits from the possible altruism of a large population. Fig.\ref{fig:Fig-3}A shows when cooperation emerges, solving $rB>C$ for $k_g/K$ as a function of $p_{\bar g}$. As before, the contours define a boundary, above which more cooperation increases fitness.
We also plot this solution in Fig.~\ref{fig:Fig-3}B for $f_{1g|1g}$ and $p_{\bar g}$, which is comparable to Fig.~\ref{fig:Fig-2}B when $q=0$. 

Second, cooperation or defection could conditionally depend on the group identity of the interaction partner. Then the in-group and out-group probabilities of cooperation $p_{gi|g'}$ evolve in opposite directions so that $p(gi|g) \rightarrow 1$, and $p(gi|g'\neq g) \rightarrow 0$. This may still mean that the average cooperative behavior $p_{gi} = \frac{k_g}{K} p(gi|g) + \frac{K- k_g}{K} p(gi|g'\neq g)$ is small if most interactions are external, as illustrated in Fig.\ref{fig:Fig-3}A, but cooperation will now be stabilized within each group. We note in Fig.\ref{fig:Fig-3}B that this need not be the final situation. Coordination and cooperation can continue to spread across groups selectively, as larger sets of group markers form new collective identities associated with in-group behavior through the mechanisms described in Table \ref{tab:Table_1}D. In this way, pedigree relations (kin selection) may often be the most natural basis for in-group formation~\cite{Hamilton1964,Queller1992}, but mutualistic interactions can also spread if they become associated with recognizable collective signals and beneficial phenotypic behaviors. Larger scale cooperative behavior, such as in human cultures, can then evolve, enabled by the expansion of group markers in tandem with population structuring.

\section*{Discussion}
The equivalence between the replicator equation and Bayes' theorem has recently been noted by several authors~\cite{harper2009replicator,Campbell2016,suchow_evolution_2017,Czegel2022,czegel_multilevel_2019}. However, this connection has remained largely a mathematical curiosity. Here, we showed how to operationalize this equivalence as a constructive and analytical tool in evolutionary population dynamics. Writing fitness as $w_s(e) = \alpha(e)\,p(e|s)$ identifies it as a conditional probability model of the environment, a pre-selection property of the phenotype rather than a record of its reproductive outcome. This resolves the circularity of the standard definition~\cite{Millstein2016,Mills1979}, while preserving the central principle of evolution:  Fitter types, with more predictive models of the environment~\cite{hledik_accumulation_2022}, are amplified by selection over time. Thus, this Bayesian redefinition of fitness  leads to the expected faster population growth of better adapted lineages but also gives these dynamics an explicit probabilistic and causal grounding.

We showed that this conceptual shift has three productive consequences. First, any conditional distribution $p(e|s)$ over the environment defines a valid fitness function, providing a direct recipe for model construction. The normalization factor $\alpha(e)$ separates the absolute growth rate of the population from the relative fitness that drives selection. Second, the long-run average growth rate (log-fitness) — the quantity maximized by selection over probabilistic environments — becomes the mutual information $I(E,S)$ between population structure and environmental statistics. This identifies information maximization as the fundamental principle of natural selection~\cite{frank2009natural,bergstrom_shannon_2004}. Additionally, it shows that additive (linear) contributions to fitness arise only when phenotypic factors are statistically independent, a condition that holds in log-fitness (growth rate) but not in fitness itself. The weak-selection approximation of population genetics recovers this linearity as a first-order expansion. Third, log-fitness naturally takes a polynomial form in phenotypic variables. Sign changes in its coefficients, as environmental parameters vary, mark phase transitions between qualitatively different selected states. In evolutionary game theory, this polynomial structure yields the threshold conditions for cooperation to emerge, recovering and extending Hamilton-type rules. It also identifies environmental probabilities as control parameters, analogous to temperature or pressure in physical systems. 

This last point also unifies distinct mechanisms proposed for the emergence of cooperation. We showed that the condition for cooperation in the Prisoner's Dilemma is that the regression $r = f_{1|1} - f_{1|0}$ be nonzero. Statistical independence ($r=0$) is exactly the absence of information between agents. Because mutual information vanishes if and only if a conditional distribution equals its marginal, the cooperation condition is, in fact, that $ p(s_j|s_i) \neq p(s_j)$, which also implies $I(S_1,S_2) > 0$. The many diverse mechanisms invoked to explain cooperation --- kin selection~\cite{Hamilton1964,Queller1992}, direct reciprocity and punishment~\cite{Trivers1971,Axelrod1980,sigmund2001reward,nakamaru2009runaway,rand2010anti}, spatial and network structure~\cite{nowak2004emergence,allen2014games,chen2015competition}, signaling~\cite{frank1987homo} -- emerge from this unifying perspective as alternative means for generating mutual information within populations. Note also that the well-mixed replicator equation of EGT forecloses the emergence of cooperation~\cite{cressmanTao2014} precisely
because it sets this information to zero by construction, see Supplementary Information. Our three applications should now also emerge as facets of a single unifying process: selection maximizes the mutual information between population structure and its environment, whether that environment is exogenous (task switching, $I(E,S)$), endogenous and co-evolving (games, $I(S_1,S_2)$), or organized by group identity.

The Bayesian-replicator mapping is exact in the setting considered here: pure selection on asexual haploid replicators in discrete time. Other evolutionary forces modify these conclusions in tractable ways. Mutation corresponds to adding a Markov transition operator to the Bayesian update, extending the inference interpretation to hidden Markov model filtering~\cite{akyildiz_replicator_2017,pathiraja_wacker_2025}. Genetic drift prevents full optimization of mutual information, setting an upper bound on the information that selection can accumulate in small populations~\cite{hledik_accumulation_2022}. Maximum-entropy approaches to quantitative trait evolution~\cite{barton_statistical_2009,bodova_general_2016,iwasa_free_1988} provide complementary tools when these forces are present. The long-run averaging that connects selection to mutual information also assumes environmental stationarity. In non-stationary environments, the appropriate quantity is a piecewise time-averaged growth rate; the optimal type may then slowly shift as the environment changes. These extensions do not undermine the likelihood definition of fitness but add complexity to the phase space of evolutionary outcomes and dynamics.

This work joins a broad effort to make information more central to evolutionary theory~\cite{kimura_natural_1961,frank2009natural,bergstrom_shannon_2004,rivoire2011value,tkavcik2016information}. Our contribution relative to this growing literature is the explicit constructive path from phenotype-environment models to information quantities as averages of log-fitness, and the demonstration that this path yields concrete analytical results — phase diagrams and Hamilton-type rules — that are otherwise difficult to derive in a unified way. The connection extends naturally to the computational perspective on evolution: selection under sexual recombination implements a multiplicative weights update algorithm~\cite{chastain_algorithms_2014,livnat_papadimitriou_2016}, and Darwinian replicator dynamics can emerge as an exaptation from selection for accurate Bayesian computation~\cite{csillag_bayes_darwin_2025}. Together, these results suggest that evolution, Bayesian inference, and statistical learning share a common computational substrate. The extent to which these communalities can become more productive in ecology and evolution should continue to be a focus of future research.  

Several directions for research follow from these results. Empirically, fitness defined via $p(e|s)$ points toward measuring the conditional relationship between phenotypes and environmental states — identifying which environmental dimensions have fitness consequences and estimating likelihood functions directly. Theoretically, the log-fitness (polynomial) framework provides a general foundation for analyzing phase transitions in evolutionary dynamics. A classification of evolutionary transitions - for example, first- or second-order - and their analytical structure in finite populations is  a subject of great interest for characterizing tipping points versus gradual change in natural history.  The structure of conditional dependence among phenotypes, analyzable through probabilistic graphical models~\cite{koller2009probabilistic}, encodes the functional organization of groups and may help illuminate the informational basis of major evolutionary transitions~\cite{szathmary_major_1995,jablonka_lamb_2006}. More broadly, the conditions for evolvability — the ability to generalize adaptations to new environments — may be understood through principles analogous to those that govern generalization in other learning systems~\cite{kouvaris_evolution_2017,watson_how_2016,frank_generalization_2026}. All this suggests that the relationship between evolutionary biology and statistical learning may go well beyond formal analogies: They are likely to have much more to teach each other also in terms of mechanisms and dynamics.

\section*{Methods}

Python code for generating Figures~\ref{fig:Fig-2}-\ref{fig:Fig-3} is available online at https://zenodo.org/records/15003106.

\bibliography{main}

\section*{Acknowledgements}
We thank Stefano Allesina, Linn\'ea Gyllingberg, David Krakauer, Michael Lachner, and Ruy Ribeiro for useful discussions and Duncan Rocha for feedback on the figures. This work is supported by the Conard Initiative on the Foundations of Complexity (to L.M.A.B), the Institute for New Economic Thinking at the Oxford Martin School (to J.T.K), and the National Science Foundation Graduate Research Fellowship Program (Grant No. DGE 2140001 to B.J.G.)

\section*{Author contributions statement}

L.M.A.B. conceived the research and developed the main theory. L.M.A.B. and J.T.K. developed the numerical experiments and conducted the analysis. L.M.A.B., B.J.G and J.T.K. wrote the manuscript. J.T.K. produced the figures. All authors reviewed the manuscript.


\newpage
\renewcommand{\thefigure}{S\arabic{figure}}
\setcounter{figure}{0}
\renewcommand\thetable{S\arabic{table}}  
\setcounter{table}{0} 
\renewcommand{\theequation}{S.\arabic{equation}}
\setcounter{equation}{0} 
\section*{Supplementary Information}

\subsection*{Average growth rates and Information}
Here, we give additional details about the connection between population growth rates and information. In Eq.~\ref{eq:information} we wrote the log-fitness, and equivalently the instantaneous growth rate, as 
\begin{eqnarray}
  \gamma_s \tau = \ln w_{s}(e) = \ln \alpha(e) + \ln p(e|s)= \ln \alpha(e) + \ln p(e) + \ln \frac{p(e|s)}{p(e)}. 
\end{eqnarray}
Below, we take $\tau=1$ for simplicity. The population dynamics expose these instantaneous quantities to a 
sequence of states of the environment, resulting in averaging over $p(e)$. This is due to the law of large numbers
\begin{eqnarray}
{\bar \gamma}_s = \frac{1}{T} \sum_{t=1}^T \ln \big[\alpha(e_t) p(e_{t}|s)\big] \simeq \ln \alpha + \sum_e p(e) \ln p(e) + \sum_e p(e) \ln \frac{p(e|s)}{p(e)}.
\end{eqnarray}
which is also the average log-fitness.  We can write these averages in terms of information quantities, leading to
\begin{eqnarray}
{\bar \gamma}_s = \ln \alpha - H(E) - D\big(p(E)\|p(E|s)\big).
\end{eqnarray}
Here $\ln \alpha= \sum_e p(e) \ln \alpha(e)$, $H(E) = - \sum_e p(e) \ln p(e)$ is the entropy of the environment, and the $D\big(p(E)\|p(E|s)\big)\geq 0$ is the relative entropy or Kullback-Leibler divergence, which is non-negative.  It follows that the average growth rate is maximized when $D$ is minimized at $p(e|s)=p(e)$. This means that type $s$ has a perfect statistical model of its environment. The other two terms are also interesting but do not affect the relative fitness because they are independent of $s$. They show that the magnitude of the payoffs $\ln \alpha > H(E)$ for there to be growth, and that environments of decreasing complexity, $H(E)\rightarrow 0$, are more easily predicted and yield higher average rewards. These results are known in statistical models of gambling and bet-hedging.

Finally, we note that when we average the $s$-type growth rates over the new state of the population (posterior) we obtain
\begin{eqnarray}
\sum_{s,e} f(s|e) p(e) \ln w_s = \ln \alpha - H(E) + \sum_{s,e} f(s|e) p(e) \ln \frac{p(e|s)}{p(e)}= \ln \alpha - H(E) + I(E,S),
\end{eqnarray}
where $I(E,S)$ is the (mutual) information between the population type frequencies and their environment. This shows that maximal population growth across types corresponds to the population structure that maximizes the mutual information $I(E,S)$, as indicated in Figure~\ref{fig:Fig-1}.

\subsection*{Analysis of the Bayesian Task Switching Problem}
Here we show the explicit Bayesian iteration for the task switching problem of Table~\ref{tab:Table_1}A-B, confirming that $f(s)\to p(e)$ in the long run. Because the problem is symmetric, it only matters whether the task is aligned with the environment ($s=e$) or not. We can see this by explicit inspection 
\begin{eqnarray}
  f(s=1|e)= \frac{p(e|s=1)}{{\hat p}(e)} f(s=1) = \frac{p(e|s=1)}{p(e|s=1) f(s=1) + p(e|s=0)f(s=0)} f(s=1).
\end{eqnarray}
Considering the two possible states of the environment and using Table~\ref{tab:Table_1} leads to
\begin{eqnarray}
f(s=1|e=1) = \frac{p(e=1|s=1)}{p(e=1|s=1)f(s=1) + p(e=1|s=0)f(s=0)} f(s=1) = \frac{4}{1+3f(s=1)} f(s=1).
\end{eqnarray}
Because the relative fitness $\frac{4}{1+3f(s=1)}\geq 1$, the conditional posteriors will converge to unity. Conversely, the unaligned probability
\begin{eqnarray}
f(s=1|e=0) = \frac{p(e=0|s=1)}{p(e=0|s=1)f(s=1) + p(e=0|s=0)f(s=0)} f(s=1) = \frac{1}{4-3f(s=1)} f(s=1).
\end{eqnarray}
Now, the relative fitness $\frac{1}{4-3f(s=1)}\leq 1$, so that the posterior probability always reduces. By symmetry, the results hold when $s=0$. Thus, the long-term limit of the dynamics is $f(s=e|e)=1, \ f(s\neq e|e)=0$. This result enables us to determine the equilibrium state of the marginal, $f(s)$
\begin{eqnarray}
  f(s) = \sum_e f(s|e)\,p(e) \;\xrightarrow{\text{long run}}\; p(e=s).
\end{eqnarray}
which means that the task is fully aligned with the environment, in probability. This result also follows directly from the interpretation of the average growth rate as information and its maximization, as discussed in the main text.

\subsection*{The standard EGT replicator equation is a mean-field approximation}

The most common starting point in evolutionary game theory (EGT) is the
continuous-time "replicator equation" for the frequency $f_i$ of type
$i$ in the population,  written as
\begin{equation}
  \frac{d f_i}{dt} \;=\; \bigl(\phi_i -{\phi}\bigr) \,f_i,
  \qquad
  \phi_i \;=\; \sum_j A_{ij}\,f_j,
  \qquad
  {\phi} \;=\; \sum_i \phi_i\,f_i.
  \label{eq:replicator_std}
\end{equation}
We now show that this is a mean approximation to the general Eq.~\ref{eq:frequency-update}. 
Here, the matrix $A_{ij}$ encodes the game structure. Its elements are the payoffs  received by a type-$i$ individual interacting with another of  type-$j$. 
Then, the quantity $\phi_i = \sum_j A_{ij} f_j$, is the "expected" payoff of type $i$, the  average over all opponent types weighed by their frequencies. (Note that $\phi_i$ is sometimes referred to as "fitness" in EGT; we dispel this potential confusion below).    ${\phi} = \sum_i \phi_i f_i$ is the mean payoff, i.e.\ the weighted average of $\phi_i$ over all types.  

The quantity $\phi_i - \phi$ is the relative (Malthusian) per capita growth rate. Types with $\phi_i > \phi$ grow in frequency, those with $\phi_i < \phi$ decline. As in the main text, the fitness is $w_i = e^{\phi_i}$, so that $\phi_i = \ln w_i = \gamma_i$ ($\tau=1$). The EGT equation is therefore written in terms of growth rates, which is also why the entries $A_{ij}$ may be positive or negative. We now show that this equation arises from the per-interaction dynamics of the main text as an infinite-population, time-averaged (mean-field) limit, corresponding specifically to the statistically independent ($r=0$) row of Table~\ref{tab:Table_1}E.

\subsubsection*{Exact interaction-by-interaction dynamics}

We now derive the mean-field EGT equation, Eq.~\ref{eq:replicator_std},  as a particular case of the general Bayesian replicator dynamics developed in
the main text.  Since real interactions are episodic, the key is to start with discrete time steps, when a focal individual
of type $s_i$ meets a \emph{single} interaction partner of type $s_j$.  The focal individual's fitness is
\begin{equation}
  w_{s_i}(s_j) \;=\; \alpha \,p(s_j \mid s_i),
  \label{eq:exact_fitness}
\end{equation}
where $p(s_j \mid s_i)$ is the likelihood that type $s_i$ assigns to
the "environmental state" $e = s_j$, and $\alpha$ is the normalization factor
of the main text.  The resulting frequency update is the exact
Bayesian replicator equation (Eq.~2 of the main text):
\begin{equation}
  f_{s_i}(t) \;=\; \frac{w_{s_i}(s_j)}{{w}(s_j)}\,f_{s_i}(t-1).
  \label{eq:stochastic_update}
\end{equation}
Because encounters with $s_j$ are probabilistic, this update is \emph{stochastic}:
a different opponent is met each time step, so the fitness
$w_{s_i}(s_j)$ is a random variable.  No approximation so far has been made.

Now, the standard EGT replicator equation replaces this stochastic,
interaction-by-interaction dynamics with a \emph{deterministic} update
driven by the average growth rate for type $i$.  To see how this arises,
note that the growth rate of type $s_i$ from a single
interaction with type $s_j$ is
\begin{equation}
  \gamma_{s_i}(s_j) \;=\; \ln w_{s_i}(s_j)
                    \;=\; \ln \alpha \;+\; \underbrace{\ln p(s_j\mid s_i)}_{\displaystyle A_{ij}},
  \label{eq:single_gamma}
\end{equation}
where we identify $A_{ij} \equiv \ln p(s_j \mid s_i)$ as the
$(i,j)$-entry of the EGT payoff matrix.

Now suppose type $s_i$ undergoes $T$ interactions in the time interval $\tau$, with opponents
drawn i.i.d.\ from the population frequencies $\{f_j\}$.  By the law of large numbers, the averaged growth rate converges
exactly as $T \to \infty$:
\begin{equation}
  {\gamma}_{i}
  \;=\; \frac{1}{T}\sum_{t=1}^{T} \ln w_{s_i}(s_j^{(t)})
  \;\xrightarrow{\;T\to\infty\;}
  \; \mathbb{E}_{f}\!\bigl[\ln w_{s_i}(s_j)\bigr]
  \;=\; \sum_j A_{ij} \ f_j\
  \;=\; (A{\bf f})_i
  \;\equiv\; \phi_i.
  \label{eq:mf_limit}
\end{equation}
This is the standard EGT payoff, $\phi_i = \sum_j A_{ij} f_j$.
The deterministic replicator equation, Eq.~\ref{eq:replicator_std}  is therefore the infinite-population-averaged limit of
the exact stochastic dynamics~\ref{eq:stochastic_update}.  This becomes exact only in the $N, T \to \infty$
limit, which cannot be satisfied in general for finite (and small) time step, $\tau$.  
As a result, such approximation leads to a number of critical limitations, to which we now turn.

\subsubsection*{General consequences for the emergence of cooperation}

The key approximation in the standard EGT, Eq. ~\ref{eq:replicator_std}, of drawing interaction  partners with probability $f_j$, independently of the focal type, is precisely that $p(s_j|s_i) = p(s_j)=f_j$. This means that the conditional and marginal probabilities coincide, $f_{1|1}=f_{1|0}$, and the assortment (relatedness) necessary for the emergence of cooperation vanishes, $r = f_{1|1}-f_{1|0}=0$. The linearity of $\phi_i$ in $\{f_j\}$ is thus equivalent to the absence of correlation between partners' strategies: the mean-field replicator equation is the statistically independent ($r=0$) case of Table~\ref{tab:Table_1}E, for which cooperation never emerges in the Prisoner's Dilemma. [This is distinct from the well-mixed row ($p_1=p_2=p$, $r_{\rm m}=1$), in which the population effectively plays itself; that assumption is more favorable to cooperation and is not the limit recovered here.]

This identification also clarifies why standard generalizations of the EGT replicator equation do not, by themselves, escape $r=0$. For example, population games (with $\phi_i$ nonlinear in $f$), multiplayer games, continuous strategy spaces, and asymmetric games all retain random matching, so the partner's type remains statistically independent of the focal type~\cite{cressmanTao2014}.  Thus, relatedness ($r>0$) requires breaking random matching --- through spatial structure, kinship, repeated adaptive interactions, or group structure (main text).

In the opposite direction, the general development of statistical dependences to maximize fitness correspond precisely to the principle of maximizing information. In this light, we can now see that the diverse mechanisms invoked to explain the emergence of cooperation --- kin selection~\cite{Hamilton1964,Queller1992}, direct reciprocity and punishment~\cite{Trivers1971,Axelrod1980,sigmund2001reward,nakamaru2009runaway,rand2010anti}, spatial and network structure~\cite{nowak2004emergence,allen2014games,chen2015competition}, signaling~\cite{frank1987homo} --- are not separate theories but distinct channels for generating the same quantity: a conditional $p(s_j|s_i)$ that departs from its marginal, i.e.\ mutual information between interacting phenotypes. The well-mixed replicator equation forecloses cooperation precisely because it sets this information to zero by construction.

\subsection*{Fitness of a general dyadic game}
Here we show that any 2x2 game can be written in the form of Table~\ref{tab:Table_1}. Consider the payoff matrix $Y_{ij}$, where the first index corresponds to the focal agent and the second to the opponent. Each player has two states $0,1$. Then 

\begin{eqnarray}
w_{s_1} (s_2) = \alpha Y_{00}^{(1-s_1)(1-s_2)} Y_{01}^{(1-s_1)s_2} Y_{10}^{s_1(1-s_2)} Y_{11}^{s_1 s_2} ,
\end{eqnarray}
where $w_{s_1}(s_2)$ represents the fitness of a player in state $s_1$ when the opponent is in state $s_2$. Thus, the log-fitness can be written as 
\begin{eqnarray}
 \ln w_{s_1} (s_2) = \ln Y_{00} + \ln \frac{Y_{10}}{Y_{00}} s_1+ \ln \frac{Y_{01}}{Y_{00}} s_2 + \ln \frac{Y_{00} Y_{11}}{Y_{01} Y_{10}} s_1 s_2. 
 \end{eqnarray}
This quantity is always a polynomial in the stochastic variables $s_1,s_2$. Its polynomial coefficients will be positive or negative depending on the specific game structure, leading to different phase diagrams as shown in the main text. Averaging the log-fitness is simple. It consists of taking expectation values of each variable and their products.

\subsection*{Analysis of selection in groups}
Here we provide the detailed derivation of the cooperation condition of Table~\ref{tab:Table_1}E, continuing from the additive log-fitness of the main text ($\ln w_{g,i} = \ln \alpha + \sum_{g',j} \ln p(s_{g'j}|s_{gi})$). An intermediate situation is when the group structure introduces conditional independence across groups but not within them. In this case, we have 
\begin{eqnarray}
w_{g i} = \alpha p( \{ s_{g' j} \} |s_{g i})= \alpha \prod_{g'} p(g'| s_{g i}), \ {\rm with} \ p(g' | s_{g i} )= \prod_{j \in g'} p( \{ s_{g' j} \} | s_{g i} ) \rightarrow \ln w_{g i} = \ln \alpha + \sum_{g'} \ln w_{gi} (g'),
\label{eq:groups}
\end{eqnarray}
with $\ln w_{gi} (g')=\sum_{j \in g'} \ln p( s_{g' j} | s_{g i} )$ the group log-fitness, obtained by summing the log-likelihoods over all members of group $g^\prime$. 

If we take $\ln p( s_{g'j} |s_{gi}) = a - c s_{gi} + b s_{g' j} + d s_{gi} s_{g'j}$, generalizing the 2-agent expression in the main text, we can write the log-fitness in terms of the sums of neighbors to our focal node. In general, these sums depend on the group structure and specifically on how interactions vary within and across groups. The relative size of these groups then becomes an important factor. We take the size of group $g$ to be $k_g$, and the total number of connections (neighbors) to be $K=\sum_g k_g = N-1$, where $N$ 
is the total population size across all groups $N=\sum_g N_g$, and $-1$ subtracts self-interactions. Many rich and complex situations can be explored as these possibilities of size and interaction structure are varied. Within our assumptions, we can then write the log-fitness as
\begin{eqnarray}
\ln w_{g i} = a K - c K s_{g i} + b \sum_{g' ,j} s_{g'j} + d s_{g i}\sum_{g' ,j} {s_{g' j}}.  
\label{eq:network}
\end{eqnarray}
We see that Eq.~\ref{eq:network} is identical to Eq.~\ref{eq:basic}, as
\begin{eqnarray}
 \frac{1}{K} \ln w_{gi} = a -c p_{gi} +b p_2 + d p_{gi} p_2
\end{eqnarray}
with the identification $p_{gi}=p(s_{gi}=1)$ and
\begin{eqnarray}
p_2 = \frac{1}{K} \sum_{g' j} p ( s_{g'j}=1 ) =  \frac{1}{K} \sum_{g'} k_{g'} p_{g'}, \qquad p_{g'}= \frac{1}{k_{g'}} \sum_{j \in g'} p(s_{g'j}=1). 
\end{eqnarray}
The conditions for the emergence of cooperation can now be assessed within and across groups using the results in the main text. First, the agent should defect unconditionally when the other agents are independent of itself. In the well-mixed population, when all probabilities of cooperation are the same, $p_{gi}=p_2 = p$, Eq.~\ref{eq-random-mixing} still applies and a phase transition occurs at $p=p_{\min}=(c-b)/(2d)$, if $b<c$. 

The situation is more interesting when states are not statistically independent and reciprocity and exploitation become group-specific. We write the generalization of Eq.~\ref{eq-cond-polynomial} with 
\begin{eqnarray}
p_2= \frac{1}{K} \sum_{g' j} \left[ f ( s_{g'j}=1 | s_{gi} =0) +r_{g' j} p_{gi} \right], \quad r_{g' j} = f ( s_{g'j}=1 | s_{gi} =1)-f ( s_{g'j}=1 | s_{gi} =0). 
\end{eqnarray}
Note that, in the statistical independence limit $r_{g' j} \rightarrow 0$ and the probabilities become unconditional. 
By averaging these conditional probabilities over the members of each out-group
\begin{eqnarray}
    f_{1g'|1g} = \frac{1}{k_{g'}} \sum_{j \in g'} f ( s_{g'j}=1 | s_{gi} =1), \  f_{1g'|0g} = \frac{1}{k_{g'}} \sum_{j \in g'} f ( s_{g'j}=1 | s_{gi} =0),
\end{eqnarray}
we can write the fitness in terms of the reciprocity with each group, with $ r_{g'} = f_{1g' |1g} - f_{1g' | 0g}$, as
\begin{eqnarray}
p_2= \frac{1}{K} \sum_{g'} k_{g'} \left[ f_{1g'|0g } +r_{g'} p_{gi} \right].
\end{eqnarray}
We can repeat this again, by averaging across all out-groups
\begin{eqnarray}
    f_{1|1g} = \sum_{g' } \frac{k_{g'}}{K}  f_{1g'|1g}, \quad  f_{1|0g} = \sum_{g'} \frac{k_{g'}}{K} f_{1g'|0g},
\end{eqnarray}
with the reciprocity defined over the average across all out-groups, $r = f_{1|1g} - f_{1|0g}$.  Finally, we can write the fitness simply in terms of interactions with the average out-group 
\begin{eqnarray}
p_2= f_{1|0g } +r p_{gi}.
\end{eqnarray}

Then, the log-fitness can be written compactly as
\begin{eqnarray}
\frac{1} {K} \ln w_{g i} = a + b f_{1|0g} + [ r b +d f_{1|1g}  - c ] p_{gi}.
\end{eqnarray}
This equation is again similar to what we obtained above, but with population definitions of $r$, $f_{1|1g}, f_{1|0g}$. 
The interesting new possibility is that the agents' behaviors are now different within and across groups in the sense that reciprocity $f_{1g'|1g} < f_{1g|1g}$, and exploitation 
$f_{1g'|0g} > f_{1g|0g}$, for $g' \neq g$. This can lead to a variety of outcomes. However, to finish the analysis here, we consider the situation when states across groups are statistically independent. Then $r_{g'}=0$ for $g' \neq g$. This leads to two different contributions to the log-fitness 
\begin{eqnarray}
 \frac{1}{K} \ln w_{g i} = a + \underbrace{ \frac{k_g}{K} \left( b r_g + d f_{1g|1g} -c  \right)}_{\rm in-group} p_{gi} + \underbrace{ \sum_ {g'\neq g} \frac{k_{g'}}{K} \left[ \left( d p_{g'} -c \right)  p_{gi} + b p_{g'}\right] }_{\rm out-groups}.
\end{eqnarray} 
This presents a conundrum: while in-group contribution drives cooperation,   out-group interactions  select for defection. In general, the out-group is larger, so cooperation may be disrupted by population interference in group dynamics. 
There are two ways out of this dilemma. To better appreciate this, we use the simplifying notation 
\begin{equation} 
\sum_{g'\neq g} \frac{k_{g'}}{K} p_{g'} = (1-\frac{k_g}{K}) p_{\bar g}.
\end{equation}
where $p_{\bar g}$ is the probability that an agent in the out-group cooperates. Note that this leads to a simpler expression 
\begin{eqnarray}
 \frac{1}{K} \ln w_{g i} = a + \frac{k_g}{K} \left( r_g b + f_{1g|1g} d - p_{\bar g } d\right) p_{gi} + p_{\bar g} d p_{gi}- p_{gi} c + (1-\frac{k_g}{K}) p_{\bar g} b,
\end{eqnarray} 
so that the condition for cooperation to emerge is now
\begin{equation}
\frac{k_g}{K} \left( r_g b + f_{1g|1g} d - p_{\bar g } d\right) + p_{\bar g} d> c, 
\end{equation}
as shown in Table \ref{tab:Table_1}E. 

We see that there are two different paths to cooperation. The first is that the agent interacts only within its in-group $k_{g}/K \rightarrow 1$. But this foregoes fitness benefits from interactions with out-group members. The second way out is to cooperate or defect conditionally on the group identity of the interaction partner. Then the in-group and out-group $p_{gi|g'}$ can adapt differentially so that $p(gi|g) \rightarrow 1$, and $p(gi|g'\neq g) \rightarrow 0$. This may still mean that the average cooperative behavior $p_{gi} = \frac{k_g}{K} p(gi|g) +  \frac{K- k_g}{K} p(gi|g'\neq g)$ is small, if most interactions are external, but cooperation will now have acquired structure that is associated with information about group identity. Gradually extending cooperative behavior to selective out-groups also provides a pathway for the growth of cooperation aided by recognizable collective signals, leading to fluid structures of intergroup cooperation and competition as shown in Figure~\ref{fig:Fig-3}.

\newpage
\subsection*{Supplementary Figures}

\begin{figure}[h]
  \centering
  \includegraphics[width=\linewidth,scale=0.6]{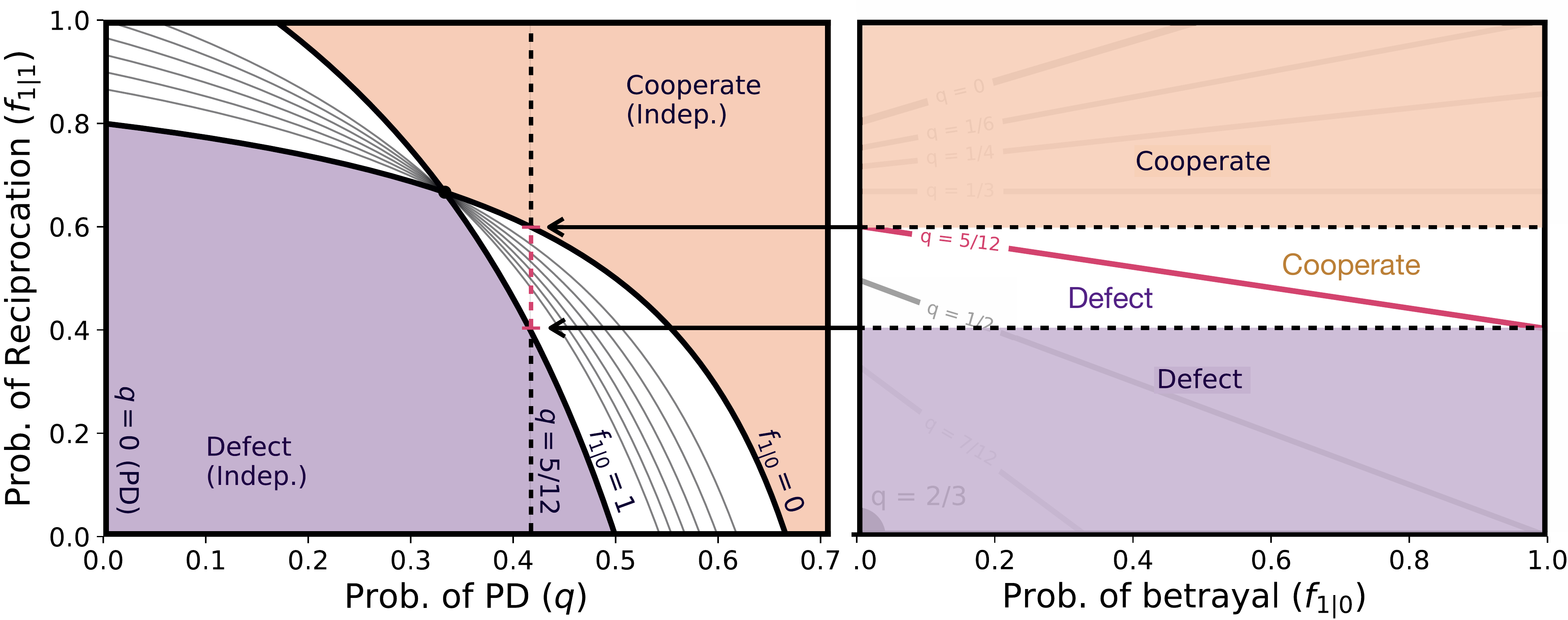}
  \caption{A guide to interpreting Fig.\ref{fig:Fig-2}B. The left diagram plots strategy boundaries parameterized by $f_{1|1},q$ along lines of constant $f_{1|0}$. Points above each trace define parameters that promote cooperation for that value of $f_{1|1}$. The shaded regions define points where cooperation or defection are promoted for all values of $f_{1|0}$. That is, the asymptotic strategies are independent of the likelihood of exploitation. In the white region, the asymptotic strategy depends on which value of $f_{1|0}$ at which the game is evaluated. The right figure illustrates these conditions along lines of constant $q$. To demonstrate how these figures complement one another, we have highlighted the results of the plot for $q=5/12$, and the slice this corresponds to on the left plot. The boundaries of the strategy zones independent of $f_{1|0}$ are matched up with dashed lines that lead to arrows, and lie entirely outside of the range of the $q=5/12$ trace. We show that in the white region, $f_{1|1},f_{1|0}$ share a linear relationship.}
  \label{fig:mixed_game_supp}
\end{figure}

\end{document}